\documentclass[copyright,creativecommons]{eptcs}
\providecommand{\event}{TFPIE 2017} 

\usepackage{graphicx}
\usepackage{subfigure}
\usepackage{listings}

\title{Functional Baby Talk: Analysis of Code Fragments from Novice Haskell Programmers}   
\author{Jeremy Singer and Blair Archibald
\institute{School of Computing Science\\
University of Glasgow\\
UK}
\email{jeremy.singer@glasgow.ac.uk \quad b.archibald.1@research.gla.ac.uk}
}
\def\titlerunning{Functional Baby Talk}
\def\authorrunning{J. Singer \& B. Archibald}

\begin{document}
\maketitle

\begin{abstract}
What kinds of mistakes are made by novice Haskell developers, as they
learn about functional programming? Is it possible to analyze these
errors in order to improve the pedagogy of Haskell? In 2016, we
delivered a massive open online course which featured an interactive code
evaluation environment. We captured and analyzed
161K interactions from 
learners.
We report typical novice developer behavior; for instance, the
mean time spent on an interactive tutorial is around eight minutes.
Although our environment
was restricted, we gain some understanding of Haskell novice 
errors. Parenthesis mismatches, lexical scoping errors and \texttt{do} block
misunderstandings are common.
Finally, we make
recommendations about how such beginner code evaluation environments
might be enhanced.
\end{abstract}




\section{Introduction}
\label{sec:intro}

The Haskell programming language \cite{hudak1992report}
has acquired a reputation for being difficult to learn.
In his presentation on the origins of Haskell
\cite{peytonjones2017escape} Peyton Jones notes
that, according to various programming language popularity metrics,
Haskell is much more frequently discussed than it is used for software
implementation. 
The xkcd comic series features a sarcastic strip on Haskell's
side-effect free property \cite{xkcd}. 
Haskell code is free from
side-effects  `because no-one will ever run it.' 
%
%

In 2016, we ran the first instance of a 
massive open online course (MOOC) at Glasgow,
providing an introduction to functional programming in Haskell.
We received many items of learner feedback that indicated
difficulty in learning Haskell. Some people found the tools
problematic:
\emph{`I have been trying almost all today to get my first tiny Haskell
program to run. The error messages on GHCi are very difficult to understand.'}
Others struggled conceptually with the language itself:
\emph{`[It]  is tedious to do absolutely anything in [Haskell] and it
made me hate Haskell with a passion.' }
Some MOOC participants have tried to learn Haskell several times:
\emph{`It's not my first attempt to learn Haskell. I always get stuck with
monad part and do-notation.'}

We want to discover how and why novice functional programmers struggle
to learn Haskell. 
What are the common mistakes they make?
Once we know the key issues, we can engage with stakeholders such
as educators, tool implementers and language designers,
to address the problems in various ways.
\begin{enumerate}
\item \textbf{Improved pedagogy}: More focused textbooks, exercises and online
  help will provide better support to aid learners to avoid
  these mistakes. This is why Allen and Moronuki \cite{allen}
  wrote a new textbook
  based on feedback from years of tutoring Haskell novices. They claim that
  `the existing Haskell learning materials were inadequate to the needs of beginners. This book developed out of our conversations and our commitment to sharing the language.'
\item \textbf{Error-Aware toolchain}: The standard Haskell tools often feature
impenetrable error messages \cite{type}. If we are aware of particular
difficulties, then we can provide dedicated support, or customized
error messages, to handle these problems. The Helium system
\cite{heeren2003helium} is motivated by such considerations.
\item \textbf{Modified language}: It may be sensible to lobby the
  Haskell language committee 
to remove incidental complexity that causes
problems in the first place. For instance, Stefik \cite{stefik2013empirical}
describes how the Quorum programming language was designed and refined based
on empirical user studies. Haskell Prime process activities are
similarly user-driven.
\end{enumerate}

But how do we identify the common mistakes that novices encounter?
Generally, we rely on first-hand anecdotal evidence.
As educators, we think back to how we, as individuals, learned
functional programming. However \emph{a sample size of one} leads
to the fallacy of \emph{hasty generalization}.
If we teach a Haskell course, then we might consider our cohorts of
students. The class sizes are relatively small---perhaps tens or at
most hundreds of students.
Another problem is lack of class diversity. The student population is
localized, has a consistent level of of education, and undergoes
a common learning experience.

We now live in the era of massive open online courses (MOOCs). 
MOOC platforms have extensive
support for data capture and learner analytics. 
A MOOC can reach a massive and diverse class, scaling to thousands of
students from a range of backgrounds.
Table \ref{tab:mooc:numbers} shows the numbers of learners
who signed up for various functional programming MOOCs. Public sources
are given where available, and the other statistics were obtained from
personal emails to the lead educators.
The data in Table \ref{tab:mooc:numbers} is presented in the same
style as Jordan \cite{jordan2014initial} who surveys general MOOC
participation trends.

\begin{table}
\begin{tabular}{p{2.0cm}|c|p{1.5cm}|p{2.0cm}|c|r||r|r}
{\small \textit{Course title}} & 
{\small \textit{Platform}} &
{\small \textit{Institute}} &
{\small \textit{Lead Educators}} &
{\small \textit{Run (year)}} &
{\small \textit{Wks}} &
{\small \textit{Signups}} &
{\small \textit{Compltns}}
  \\ \hline \hline
Functional Programming Principles in Scala & Coursera & EPFL
& Odersky, Miller
 & 1 (2012)  & 7 & 50K \cite{scalamooc} & 9.6K \\ \hline
Introduction to Functional Programming & edX & Delft & Meijer 
& 1 (2014) & 8 & 38K \cite{edxmooc}  & 2.0K \\ \hline
Introduction to Functional Programming in OCaml & FUN & Paris Diderot
                                    & Di Cosmo, Regis-Gianas, Treinen
& 1 (2015)& 6 & 3.7K & 300  \\ \hline
Functional Programming in Haskell & FutureLearn & Glasgow
& Vanderbauw-hede, Singer
& 1 (2016)& 6  & 6.4K & 900 \\ \hline
Functional Programming in Erlang & FutureLearn & Kent & Thompson
& 1 (2017) & 3 & 5.6K & 400 \\ \hline \hline
\end{tabular}
\caption{\label{tab:mooc:numbers}Summary of introductory functional programming
  MOOCs, including statistics where available or confirmed personally
  (note that completion metrics are not directly comparable across MOOC providers)}
\end{table}

We designed the learning materials in our MOOC so as to 
capture learners' interactions via an online system, particularly
their interactive programming exercises.
Our aim is to analyze the program fragments, to discover
what learners find difficult, and whether we might be able to address any of their
issues using the strategies outlined above.


The contributions of this paper are as follows:
\begin{itemize}
\item We provide an  experience report about running a functional
  programming
MOOC, from the perspective of educational practitioners. 
To the best of our knowledge, this is the first report of this kind.
\item We present empirical analysis of learner behavior, showing how 
the typical MOOC dropoff rate is observed in our interactive coding 
exercises.
\item We advocate a feedback-driven approach to
functional programming education, facilitated by
analysis of large and diverse learner communities.
\end{itemize}

The paper structure is as follows:
Section \ref{sec:platform} describes the interactive coding
environment we deployed for novice learners; 
Section \ref{sec:rqs} provides a quantitative analysis of learner
behavior, based on activity logs;
Section \ref{sec:threats} outlines threats to validity;
Section \ref{sec:relw} reviews related work; 
finally, Section \ref{sec:concl} concludes and discusses future work.


\section{Evaluation Platform}
\label{sec:platform}

\begin{figure}
\begin{center}
\includegraphics[width=0.9\textwidth]{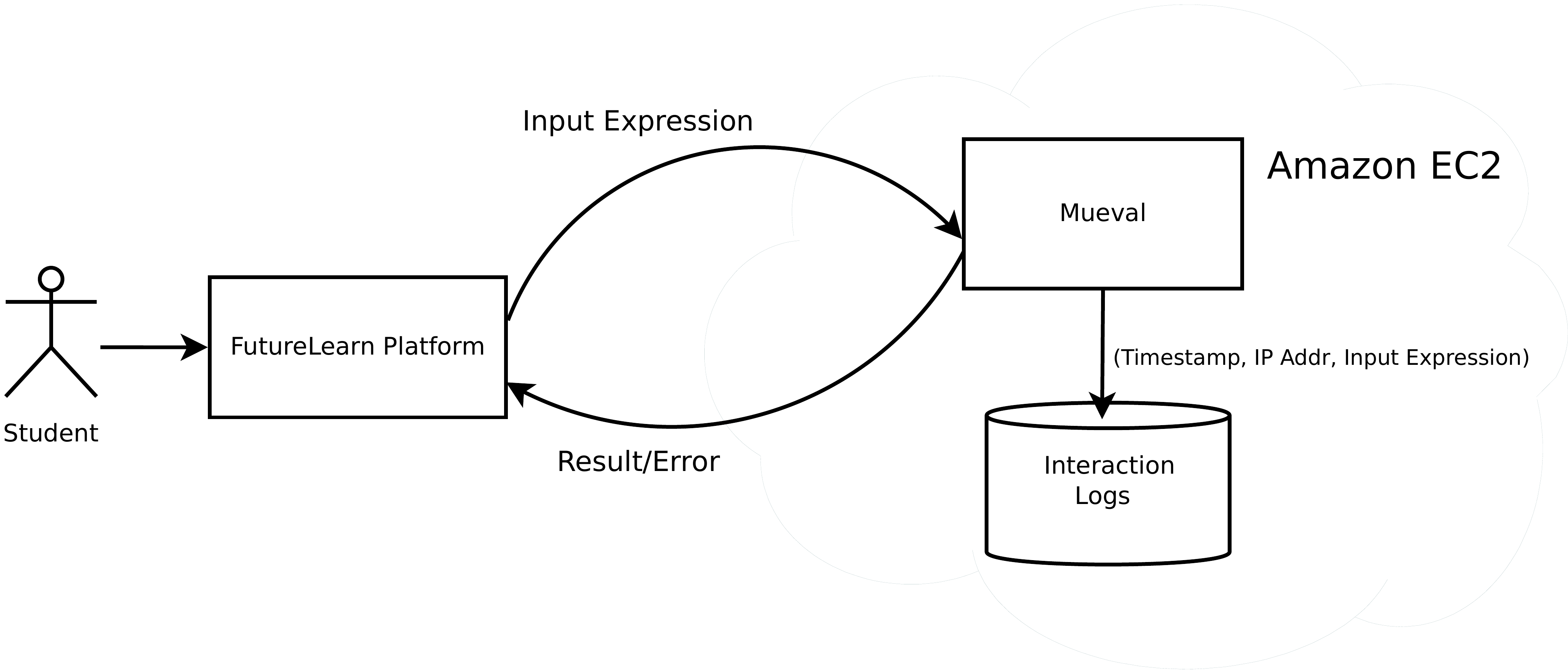}
\end{center}
\caption{\label{fig:arch}Architectural diagram of our Haskell code
  evaluation infrastructure, which was hosted on Amazon Elastic Compute Cloud
  (EC2)}
\end{figure}

Our students use a browser-based read-eval-print-loop (REPL) system
for the first three weeks of our six week course. 
This system is based on the \emph{tryhaskell} framework \cite{done} which
has an interactive Javascript front-end to parse expressions and
provide user feedback, coupled with a server-based back-end that uses
the \emph{mueval} tool \cite{mueval} to evaluate simple Haskell expressions
in a sandboxed environment. Figure \ref{fig:arch} shows the
architecture of our system. We hosted three load-balanced
mueval servers on Amazon EC2 instances for the duration of our course.

\begin{figure}
\begin{center}
\includegraphics[width=0.9\columnwidth]{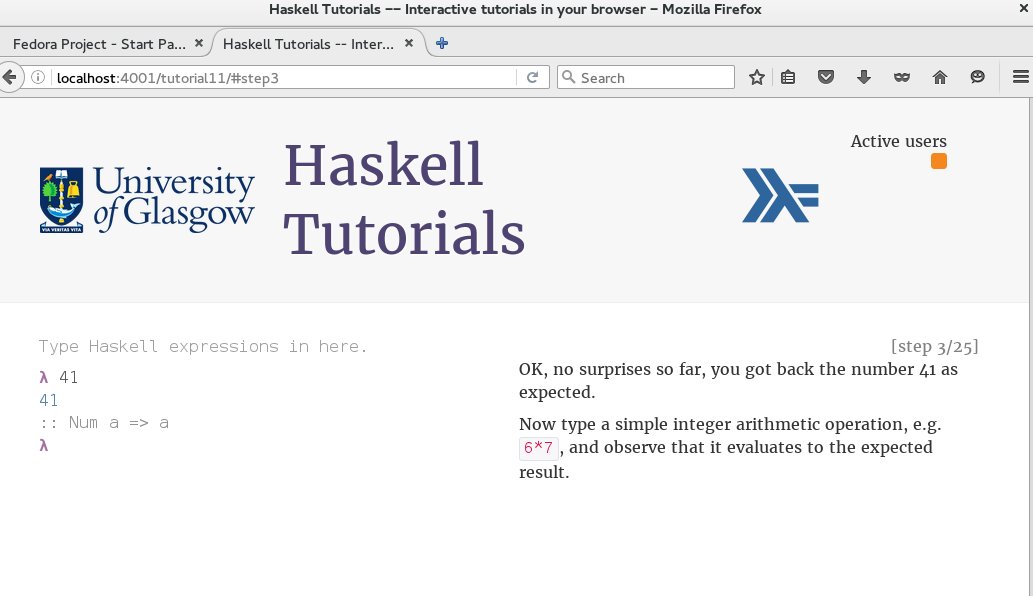}
\end{center}
\caption{\label{fig:tryhaskell}Screenshot of our interactive Haskell
  tutorial front-end, based on the tryhaskell system}

\end{figure}

At the client-side, a learner follows a series of instructions in an 
interactive prompt-based Javascript terminal, entering
one-line Haskell expressions. See Figure \ref{fig:tryhaskell} for an
example screenshot.
These expressions are sent to the server, where they are executed
by the mueval interpreter in a stateless way. The result is returned
to the client and displayed in the REPL.
The mueval interpreter logs each one-line expression it attempts to
evaluate,
along with a time stamp and an originating IP address.

We make several minor modifications to the original tryhaskell system,
which we have forked on github \cite{forkedth}.
\begin{enumerate}
\item We use Javascript to wrap \texttt{let} expressions in a context so we can emulate
persistent \emph{name bindings}. This compound \texttt{let} expression is
automatically prefixed to each one-liner that the user enters, before
the code is sent to mueval.
\item Whereas the original tryhaskell system has a fixed sequence of
  interactions, we allow a \emph{flexible scripted set of interactions}
  that highlight the particular topics we are covering at each stage
  of the MOOC.
\item We run multiple evaluation servers concurrently, behind a
  cloud-based \emph{load-balancer}. A client can be served by any evaluation
  server, since all interactions are stateless.
\end{enumerate}


\section{Learner Data Analysis}
\label{sec:rqs}

\begin{figure}
\begin{center}
\includegraphics[width=0.9\textwidth]{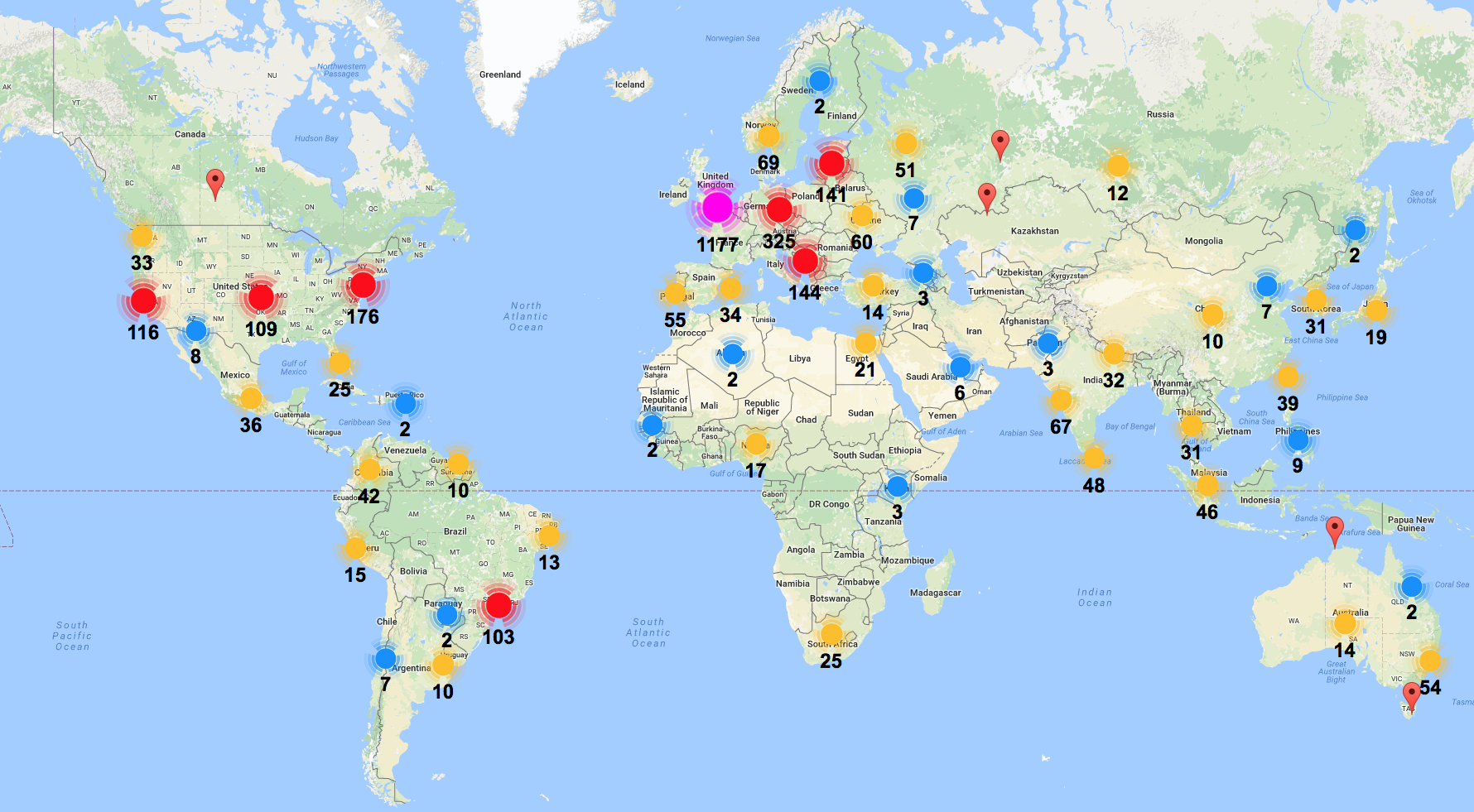}
\end{center}
\caption{\label{fig:map}Map of geo-located IP addresses that accessed our
  servers during the course. Multiple accesses from the same
  IP address are only counted once in this map. Some IP addresses
  failed geo-location.}
\end{figure}

In this section, we analyze the data from our MOOC learner population.
First we summarize the log files gathered from the three load-balanced
servers.
The data was collected from 21 September to 3 November 2016, which
included the majority of the course.
The log files record 161K lines of interactions, comprising around
17MB of data.
We logged 3.3K unique IP addresses, which broadly corresponds to
the course's 3.1K `active learners' reported by the FutureLearn platform, i.e.\
those who completed at least one step of the course.
Figure \ref{fig:map} shows aggregated counts of
these geo-located IP addresses, superimposed
on a world map.

Our log files are available on
github \cite{logs}.
We have anonymised the IP addresses by replacing each distinct IP address
with an integer value. Otherwise, the log data has not been modified.
We encourage other researchers to mine and analyse our logs.

The following sections drill down into more detailed analysis of this data.

\subsection{Interactive Sessions}
\label{sec:rqs:sessions}

Learners use our interactive tutorial platform in a session-based
manner. Each tutorial exercise is designed to take between 5 and 15
minutes, if the user reads the prompts carefully and constructs proper
Haskell expressions for evaluation. There are seven exercises in the
first three weeks of the course.

From our log files, we attempt to reconstruct these sessions.
Hage and van Keeken \cite{hage2007mining} refer to such sessions
as \textit{coherent loggings}.
We group together log entries that originate from the same IP address,
with an interval of less than 10 minutes between successive
interactions. The tutorial system is set up as a read/eval/print
loop. As a dual to this, we model the user as a write/submit/think
loop. We set a 10 minute limit on iterations round this loop.

We acknowledge that IP address might not be an entirely accurate
identifier for a learner, but it is the best proxy we can extract
from our log files. Learners were not required to authenticate when they
accessed the interactive tutorials.

\begin{figure}
\begin{center}
\includegraphics[width=10cm]{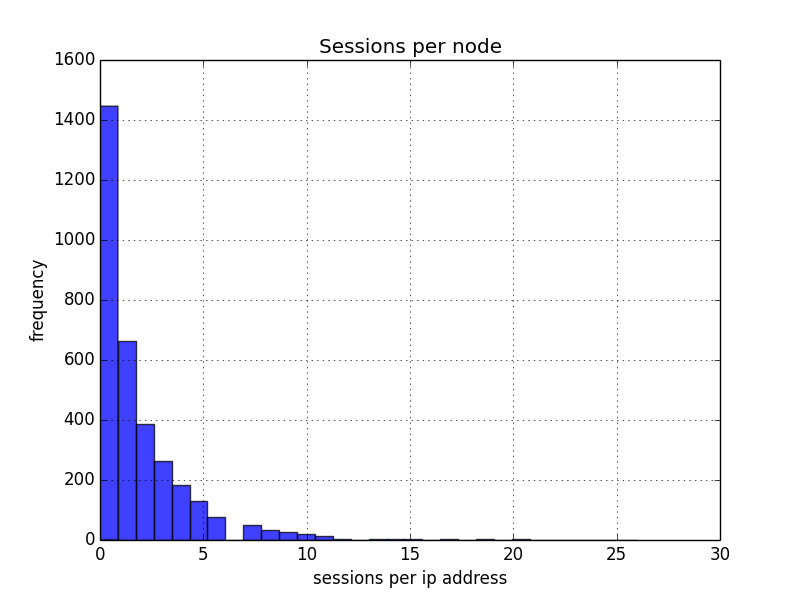}
\end{center}
\caption{\label{fig:rqs:numsessions}Histogram of sessions split into bins
  based on number of sessions per IP address}
\end{figure}

In total, we identified 5.7K sessions from our logs. The mean number
of sessions per IP address is 1.72. This data shows us that many
learners
did not complete the full set of seven tutorials.
Figure \ref{fig:rqs:numsessions} shows the number of sessions
per IP address, split into 30 equal-sized bins.

\begin{figure}
\begin{center}
\includegraphics[width=10cm]{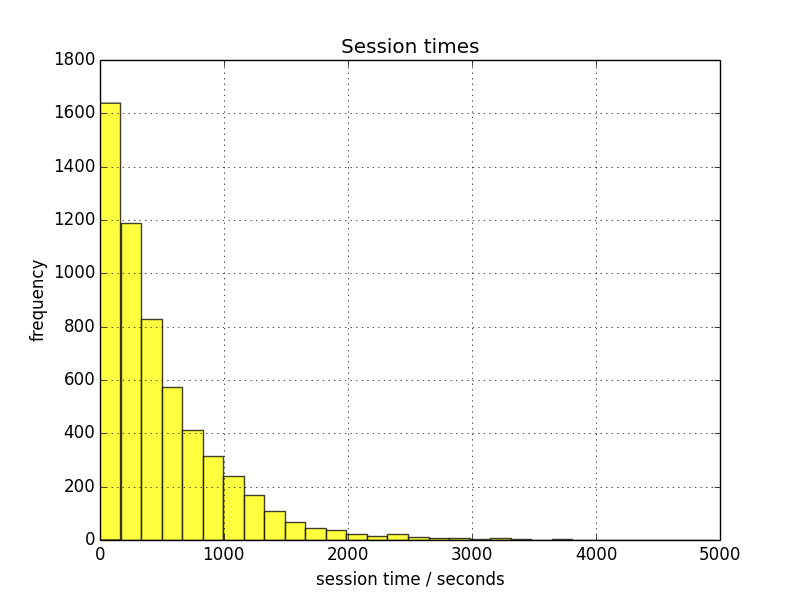}
\end{center}
\caption{\label{fig:rqs:sessiontime}Histogram of sessions split into bins
  based on the total time of each session}
\end{figure}

The mean time spent in a single session is 490s, or around 8 minutes.
This corresponds to our design intentions, and also closely follows
the observations of Guo et al.\ on MOOC video length to maximize
engagement \cite{guo2014how}.
Figure \ref{fig:rqs:sessiontime} shows the length of time per session,
split into 30 equal-sized bins.

\begin{figure}
\begin{center}
\includegraphics[width=10cm]{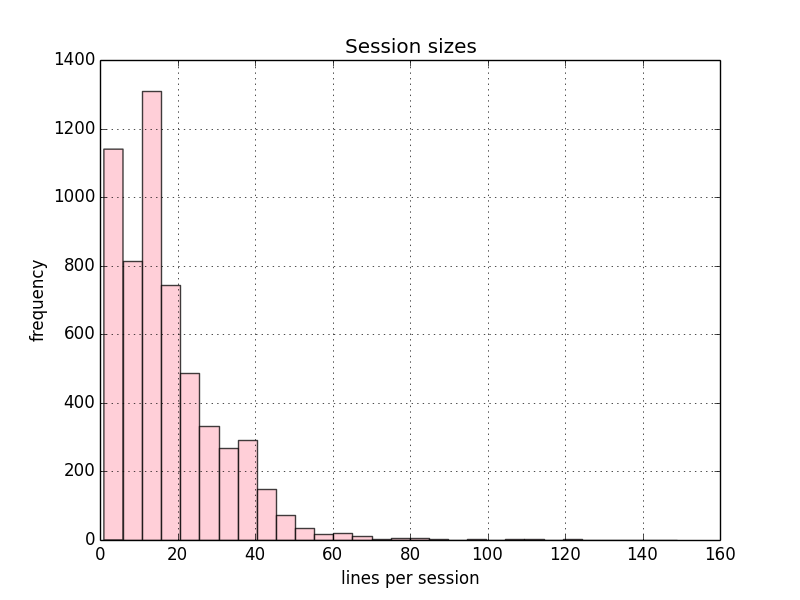}
\end{center}
\caption{\label{fig:rqs:sessionsize}Histogram of sessions split into bins
  based on the number of lines of code in each session}
\end{figure}

The mean number of Haskell expressions entered per session is 17.
Figure \ref{fig:rqs:sessionsize} shows the number of lines per
session, split into 30 equal-sized bins.

The steep drop-off apparent in these histograms is
characteristic of a fat-tailed distribution seen in 
general MOOC learner engagement, as outlined by
Clow with his theory of the \textit{funnel of participation} \cite{clow2013moocs}.

\subsection{Adventurous Coders}
\label{sec:rqs:original}

The interactive coding materials, which we host on our forked variant
of
tryhaskell,
are walk-through tutorial style exercises. Each tutorial consists of a
fixed
number of discrete steps. Each step includes some explanatory text,
then the user is prompted to enter a Haskell expression. Sometimes
we provide the correct expression directly, and the user simply copies
this.
Other times we provide hints, and users have to construct their own
expressions.
Note that we are flexible, in terms of the user entering
different code --- if the expression is at all appropriate then the
tutorial advances to the next step.

We measure the proportion of user input that is based directly
on cutting and pasting Haskell code supplied in the tutorial.
The rest of the input is modified by users, 
who have either edited the code to customize it in some way 
or written something completely different.

\begin{figure}
\begin{center}
\includegraphics[width=10cm]{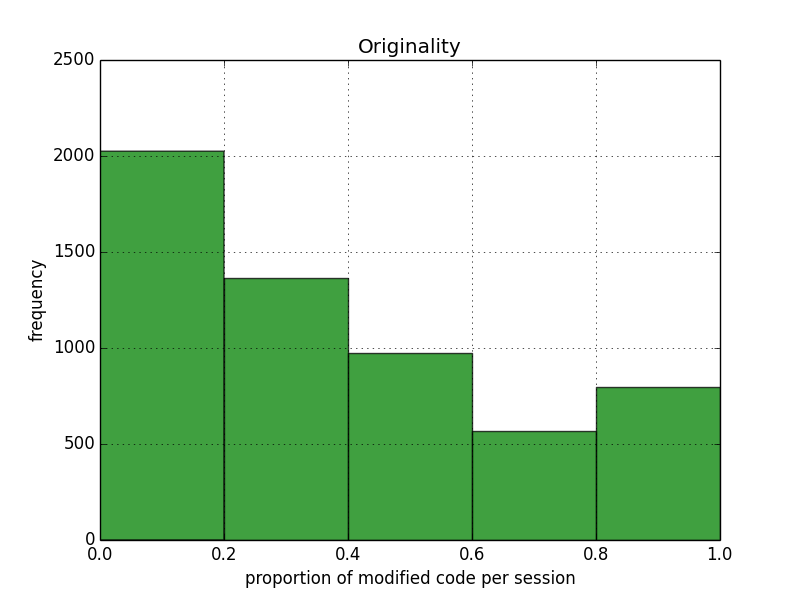}
\end{center}
\caption{\label{fig:rqs:original}Histogram of sessions split into bins
  based on proportion of modified lines of code in session}
\end{figure}

In terms of the 161K lines of user input, around 101K are unmodified lines
of code and 60K are modified.
We also analyze each session individually, to measure the proportion
of lines in each session that are modified (i.e.\ original to that
user).
Note that sessions have
different lengths, as Section \ref{sec:rqs:sessions} explains.
Figure \ref{fig:rqs:original} presents a histogram of sessions, in
terms of the proportion of original lines of code per session.
The sessions are split into five bins. We observe that the largest bin
contains sessions with minimal code modifications.

\subsection{Learner Syntax Errors}
\label{sec:rqs:errors}

We extracted the individual Haskell expressions entered by learners,
from the log file entries. We ran each of these expressions through a
Haskell parser, based on the \texttt{haskell-src-meta}
package. We discovered that 8.1K of the 161K lines could not be parsed
correctly as Haskell expressions. 

Of these unparsed lines, 2.1K are parse errors involving the
value-naming operator, \texttt{=}. In our tryhaskell REPL, we accept
variable binding operations of the form \texttt{var=expr} --- which is
not a valid expression.  A further 390 lines include unexpected
\texttt{:} characters. These appear to be attempted invocations of 
gchi commands like \texttt{:quit}, which are not supported by the REPL.

Table \ref{tab:rqs:errors} lists the most common error messages in
Haskell expressions,  as reported by our parser.
We explored these errors by conducting detailed examination of
relevant logged interactions.
We wanted to determine the high-level root cause of each error.
This is a manual task requiring significant domain expertise; we note
this does not scale particularly well.
However, we identify some apparent high-level problems, such
as:
\begin{enumerate}
\item \emph{Parenthesis mismatch}: Parenthesis characters are
  frequently unbalanced. A
  sample expression that causes this error is:
\texttt{min((max 3 4) 5))} --- 
Heeren et al.\ \cite{heeren2003helium} also state that `illegal nesting of
parentheses, braces, and brackets is a very common lexical error'.

\item \emph{Bad scoping}: Problems with \texttt{let} and \texttt{where} constructs are apparent.
The mueval expression parser does not support \texttt{where} clauses
properly. Some users had confusion with the syntax of \texttt{let}, e.g.\
\texttt{let (a,b) (10, 12) in a * 2}.

\item \emph{Misunderstanding do blocks}: Many people tried to bind values to
  names with \texttt{<-} outside a \texttt{do} block, or bind names in a \texttt{do}
  block as the final action, e.g. \texttt{do {putStrLn "nnnnn "; z <- getLine;}}

\item \emph{Complex constructs}: The mueval interpreter is particularly restrictive. It
  does not support \texttt{data} or \texttt{type} definitions, or
  definitions of multi-line functions. Several users attempted to enter
  such code, which did not parse correctly. We encouraged users to
  transition to GHCi for these constructs. Perhaps some did not read the
  supporting text, or were expert Haskell users trying to discover the
  limits of our system.

\item \emph{Incorrect syntax for \texttt{enumFromThenTo} syntactic sugar}: People
  misunderstood the \texttt{..} notation, generating incorrect list
  expressions like: \texttt{[0,1,3..10]} or \texttt{[0, 2, ..]} or
  \texttt{[1.1, 1.2, .. 2.0]} -- this may have been a problem with our
  tutorial material. Alternatively, learners may be confusing Haskell
  and Python list syntax.

\end{enumerate}

\begin{table}
\begin{center}
{\small
\begin{tabular}{p{7cm}|r}
\emph{Reported error}  &  \emph{Count} \\
\hline \hline
\lstinline!Parse error: } ! & 227 \\ \hline
\lstinline!Parse error: .. ! & 223 \\ \hline
\lstinline!Parse error: | ! & 196 \\ \hline
\lstinline!Parse error: ) ! & 174 \\ \hline
\lstinline!Parse error: , ! & 136 \\ \hline
\lstinline!Parse error: <- ! & 135 \\ \hline
\lstinline!Parse error: \\ ! & 130 \\ \hline
\lstinline!Parse error: -> ! & 114 \\ \hline
\lstinline!Parse error: in ! & 105 \\ \hline
\lstinline!Parse error: ; ! & 98 \\ \hline
\lstinline!Parse error: where ! & 91 \\ \hline
\lstinline!Parse error: ] ! & 85 \\ \hline
\lstinline!Parse error: + ! & 64 \\ \hline
\lstinline!Parse error:! Last statement in a do-block must be an expression & 63 \\ \hline
\lstinline!Parse error: else ! & 61 \\ \hline
\lstinline!Parse error: / ! & 50 \\ \hline
\lstinline!Parse error in pattern: length' ! & 45 \\ \hline
\lstinline!Parse error: virtual } ! & 41 \\ \hline
\lstinline!Parse error: let ! & 36 \\ \hline
\lstinline!Parse error: if ! & 34 \\ \hline
\lstinline!Parse error: ` ! & 33 \\ \hline
\lstinline!Parse error: { ! & 30 \\ \hline
\lstinline!Parse error: * ! & 28 \\ \hline
\lstinline!Parse error: then ! & 27 \\ \hline
\lstinline!Parse error: > ! & 27 \\ \hline
\lstinline!Parse error in pattern: l' ! & 26 \\ \hline
\lstinline!Parse error in pattern: l ! & 26 \\ \hline
\lstinline!Parse error: type ! & 25 \\ \hline
\lstinline!Parse error in pattern: f ! & 25 \\ \hline
\lstinline!Parse error in pattern:  ! & 23 \\ \hline
\lstinline!Parse error: data ! & 22 \\ \hline
\lstinline!Parse error: => ! & 21 \\ \hline
\lstinline!Parse error: - ! & 20 \\ \hline \hline
\end{tabular}
}
\end{center}
\caption{\label{tab:rqs:errors}Sorted list of most frequent parse
  errors for learner input}
\end{table}

When learners make mistakes in the programming exercises, we want
them to continue to experiment with their Haskell code to fix the
problem. The interactive tool provides context-specific feedback and
hints for common error cases. We analyze our logs to see how many
learners `give up' when they encounter an error. Of the 5.7K sessions,
3.5\% contain a Haskell expression with a parse error as the last line
of user input. This suggests a low proportion of learners abandon
sessions due to problems with errors.

As course designers, we opted to use tryhaskell because it works `out of the box' for
learners in their familiar browser environments; there is no need to
do any confusing or difficult toolchain installation before starting
to write code.
However we recognize that the tryhaskell environment, with its single
line REPL interface and lack of syntax highlighting,
is not suitable for learning anything more than the most basic Haskell expressions. 
We will need to provide more explicit signposting here, or a more
fully featured online evaluation environment. In our six week Haskell
course, we currently advise learners to transition to GHCi during the third
week, when we start to cover more advanced Haskell features that are
not supported by the REPL.

\subsection{Expression Typing Analysis}
\label{sec:rqs:types}

As before, we extracted the individual Haskell expressions entered by
learners, from the log file entries. 
We use the GHC expression evaluation facility to query the \emph{type} of
each line of learner input.
Of the lines that can be parsed, 28K lines contain expressions with
type errors. This is around 18\% of all syntactically correct expressions.

We briefly inspected the GHC type error messages to uncover common causes of
typing problems.
These include many \texttt{not in scope} naming problems. Some are
simple spelling mistakes, like \texttt{lenght},
\texttt{putsStrLn} and 
\texttt{xipwith}.
Others scoping issues are more complex, where defined names are bound
to variables with incorrect types, indicating potential
learner misunderstandings of the type system.

We detected further type mismatches via unsatisfiable type
constraints, e.g.\ the error \emph{No instance for (Num Bool) arising
  from the literal `1'} is caused by a learner attempting an equality
test between \texttt{True} and \texttt{1}.
Most of these sorts of errors derive from equality tests and if/then/else
expressions. Our interactive lessons encouraged students to explore
these facilities, so we should expect such type mismatches.


Table \ref{tab:rqs:types} lists the most frequent types observed in
learner expressions. 
We observe that \texttt{Bool}s are more popular than \texttt{Num}eric types.
Lists are fairly common. 
The frequent type constraints are generally induced by 
\texttt{enumFromTo} and
if/then/else expressions.
The majority of \texttt{IO} is due to \texttt{putStrLn} and
\texttt{getLine} which are introduced early in the course.
We see that function types are relatively infrequent. Learners can only
construct single line function definitions in our REPL environment.
Also, the first two weeks of material focus more on 
using functions (e.g.\ for list processing) instead of defining them.

\begin{table}
\begin{center}
{\small
\begin{tabular}{p{7cm}|r}
\emph{Expression Type}  & \emph{Count} \\
\hline \hline
\lstinline!Bool! & 28798 \\ \hline
\lstinline!Num a => a! & 25198 \\ \hline
\lstinline!IO ()! & 9709 \\ \hline
\lstinline!Num a => [a]! & 9701 \\ \hline
\lstinline![Char]! & 7589 \\ \hline
\lstinline!Int! & 3910 \\ \hline
\lstinline!(Num a, Ord a) => a! & 3870 \\ \hline
\lstinline!(Enum a, Num a) => [a]! & 3587 \\ \hline
\lstinline!IO String! & 2893 \\ \hline
\lstinline!Floating a => a! & 2361 \\ \hline
\lstinline!Num a => a -> a! & 1840 \\ \hline \hline
\end{tabular}
}
\end{center}
\caption{\label{tab:rqs:types}Sorted list of most frequent types
  observed for learner input expressions}
\end{table}









\section{Threats to Validity}
\label{sec:threats}

This section considers potential threats to the validity of our
findings, in terms of characterizing novice Haskell programmers.

\subsection{Lack of Generality}

The log files we analyzed were generated by learners from the first
run of a single MOOC. While the learners were drawn from varied
backgrounds, they were exposed to a single set of Haskell educational
resources on this course. 
In that sense, the findings may not be representative of a wider
variety of novices who are taught in a different way, or with
different materials.

We acknowledge the need to gather data from repeated runs of the
course, and to cross-check this data with other courses.
For instance, the Helium project \cite{heeren2003helium} reports similar data, but is at a
higher level of abstraction.

\subsection{Supported Language Subset}

The tryhaskell REPL environment, backed by the mueval utility,
handles simple one-line Haskell expression evaluation. In that sense, only a
subset of the Haskell language is supported. Hence we are limited
to exposing a subset of novice errors. 

We argue that the language features supported by mueval are
most appropriate for novice Haskell learners --- hence the errors are
still representative of those that would be experienced in the full
Haskell language. We could confirm this by replacing mueval with a
more powerful evaluation framework.

\subsection{Limited Error Analysis}

Our investigation of errors in learner input (Section
\ref{sec:rqs}) is a surface analysis. If an expression can be
parsed and type-checked, then we do not flag it as erroneous. 

We have not fully characterized the type errors, which the Helium
developers \cite{heeren2003helium} highlight as being particularly
prevalent in novice code. Our REPL does report type
errors back to learners when they initially enter the code, but we 
have not yet classified these errors in a systematic way.

Errors arise from different root causes, e.g.\ limitations of REPL and
learner misunderstandings are common causes. A deeper analysis could
identify a wider range of root causes.

\subsection{Software Incompatibility}

Some learners reported that our interactive tutorials did not work on their
machines. This may have been due to web browser incompatibilities or
OS issues. The problem seemed to occur mostly on Windows devices
(which was a popular OS with users).

Because of this, our logs might be skewed towards users with particular
browser/OS configurations, but we do not expect that this will
introduce significant bias into the results.


\section{Related Work}
\label{sec:relw}

In addition to the original tryhaskell platform \cite{done} there are 
similar online REPL systems for Scala (Scastie and Scala fiddle),
OCaml (tryOCaml) and other functional languages.
While these systems might capture user interactions, to the best of
our knowledge there is no publicly available analysis of the data.

Of the four other functional programming MOOCs listed in Table
\ref{tab:mooc:numbers}, two include integrated interactive coding
environments (Scala and OCaml MOOCs). Some survey data from the Scala MOOC
is available online \cite{scalamooc,scaladata}
although there is no published analysis of learner interactions.

The Helium system \cite{heeren2003helium}
logs the behavior of novice Haskell developers.
Hage and van Keeken \cite{hage2007mining} presents some similar
metrics and graphs to ours, based on mining data from Helium logs.
Their platform performs whole-program execution, so they can work
with a richer set of Haskell constructs.
They provide a list of common Haskell errors \cite{heeren2003helium}.
Some of their lexical errors are identical to ours, such as
parenthesis problems.
They also have details of common type errors, which we have done in a
limited way.
They capture and analyze
programs created by students at a single institution.
Their corpus comprises 30K Haskell programs, in contrast to our 161K one-line
Haskell snippets.

Thompson \cite{hugs} presents a list of common Haskell errors that are
generated in the hugs interpreter. Again, there is some overlap with
our set of errors. 

Bental \cite{bental1995why} describes an instance of the interactive 
Ceilidh framework for assignment-based ML program submission and
feedback.
She gives a categorization of 134 programming
questions submitted by students who could not generate correct ML code to
solve particular assignments. 18 of these problems are syntactic ---
which are most strongly related to the errors we have highlighted in our study.

Marceau et al.\ \cite{marceau2011measuring} 
present the DrRacket system, and how they devised a set of 
user-friendly error messages which were the outcome of systematic user trials.
They observe that parenthesis matching is a common problem in the
first lab exercise, but becomes less apparent in later labs as
students gain experience.

DrRacket \cite{findler2002drscheme}
introduces a LISP-like language to novice programmers
via a controlled sequence of gradually more
powerful language subsets.
It might be possible for us to do something similar with Haskell.
We would need to make it clear which language elements are supported
in each subset.
In effect, the tryhaskell system does define a Haskell subset, since
there are many features (e.g.\ multi-line functions and datatype
definitions) that it does not support.
It might be sensible to trigger `wait till later' warning messages if a learner tries
to evaluate anything more complex than we expect, inside our REPL.
We note that Helium \cite{heeren2003helium} and Elm
\cite{czaplicki2013async} are deliberately cut-down 
functional languages that eliminate complex concepts like
type classes.

L{\"a}mmel et al.\ \cite{lammel2013101}
advocate a systematic approach to learning Haskell, blended with 
relevant online content. They present a \emph{chrestomathy}, which 
is a highly structured set of linked resources, with an accompanying
ontology. Their approach introduces distinct programming concepts in a 
carefully controlled order.

Vihavainen et al.\ \cite{vihavainen2012multi}
describe an introductory programming MOOC that
features scaffolded tasks, which are similar to 
our Haskell interactive exercises. Their tasks include
automated tests, so online learners gain immediate
feedback. They also outline the notion of a 
learning pyramid, where various stakeholders (teachers, tutors,
more experienced learners, less experienced learners)
form relationships at different levels, in a process they term
`extreme apprenticeship'. Our FutureLearn platform facilitates
similar organic interactions between course participants, and we are
analyzing this as part of ongoing work.

Murphy et al.\ \cite{murphy17analysis} note that a small number of UK
universities (four) teach Haskell as a first programming language to Computer
Science students. This study describes general motivations for
selecting particular languages, as well as perceptions of
their relative difficulty. However the data does not give particular insights
regarding Haskell, since it is such a small proportion of the overall sample.

Large novice programmer datasets are available for other languages,
such as Java \cite{altadmri201537}.
This study analyses 18 common errors, of which the
majority are syntactic. However it also shows that semantic and type
errors are significant, and may take longer to correct.


\section{Conclusions}
\label{sec:concl}

Interactive learning environments are useful for novice developers,
but engagement is subject to the standard drop-off that characterizes
MOOC participation. We analyzed 161K lines of Haskell code submitted
for evaluation by learners, and identified some common syntactic error
patterns. Many of these are consistent with previous error
classifications reported in the literature.

Based on the errors observed in the learner code, we are targeting
specific adjustments to our course materials for the next run of our
MOOC.
Further analysis may be possible from our log files. 
For instance, we intend to perform more sophisticated
type-based analysis.
We have
published anonymized versions of our logs on github \cite{logs} for the benefit of the
research community. We will continue to record and analyze
student interactions in future iterations of our Haskell MOOC.

We feel that a richer coding tool
(featuring syntax highlighting, parenthesis matching, and
multi-line input, \textit{inter alia}) would be more appropriate to
support beginner Haskell developers in an online environment.
We are currently investigating the IHaskell kernel for the
Jupyter Notebook platform \cite{ihaskell}
as an alternative framework for scaffolded interactive exercises.

\section*{Acknowledgments}

All learner data was gathered via the FutureLearn platform.
Thanks to Wim Vanderbauwhede for co-teaching the Haskell MOOC
at Glasgow in 2016.
We acknowledge Simon Jouet, who helped with IP geo-location.
We are grateful for constructive feedback from
Anna Lito Michala, Vicki Dale, Dejice Jacob, and the
participants at the TFPIE 2017 conference.

\bibliographystyle{eptcs}
\bibliography{babytalk}

\begin{thebibliography}{10}
\providecommand{\bibitemdeclare}[2]{}
\providecommand{\surnamestart}{}
\providecommand{\surnameend}{}
\providecommand{\urlprefix}{Available at }
\providecommand{\url}[1]{\texttt{#1}}
\providecommand{\href}[2]{\texttt{#2}}
\providecommand{\urlalt}[2]{\href{#1}{#2}}
\providecommand{\doi}[1]{doi:\urlalt{http://dx.doi.org/#1}{#1}}
\providecommand{\bibinfo}[2]{#2}

\bibitemdeclare{book}{allen}
\bibitem{allen}
\bibinfo{author}{Christopher \surnamestart Allen\surnameend} \&
  \bibinfo{author}{Julie \surnamestart Moronuki\surnameend}
  (\bibinfo{year}{2017}): \emph{\bibinfo{title}{Haskell Programming from First
  Principles}}.
\newblock \bibinfo{publisher}{Gumroad (ebook)}.

\bibitemdeclare{inproceedings}{altadmri201537}
\bibitem{altadmri201537}
\bibinfo{author}{Amjad \surnamestart Altadmri\surnameend} \&
  \bibinfo{author}{Neil~C.C. \surnamestart Brown\surnameend}
  (\bibinfo{year}{2015}): \emph{\bibinfo{title}{37 Million Compilations:
  Investigating Novice Programming Mistakes in Large-Scale Student Data}}.
\newblock In: {\sl \bibinfo{booktitle}{Proceedings of the 46th ACM Technical
  Symposium on Computer Science Education}}, pp. \bibinfo{pages}{522--527},
  \doi{10.1145/2676723.2677258}.

\bibitemdeclare{inproceedings}{bental1995why}
\bibitem{bental1995why}
\bibinfo{author}{Diana \surnamestart Bental\surnameend} (\bibinfo{year}{1995}):
  \emph{\bibinfo{title}{Why doesn't my program work? requirements for automated
  analysis of novices’ computer programs}}.
\newblock In: {\sl \bibinfo{booktitle}{Proc.\ Workshop on Automated
  Understanding of (Novice) Programs, World Conference on AI and Education,
  Washington DC, USA}}.

\bibitemdeclare{misc}{mueval}
\bibitem{mueval}
\bibinfo{author}{Gwern \surnamestart Branwen\surnameend}
  (\bibinfo{year}{2014}): \emph{\bibinfo{title}{Mueval}}.
\newblock \bibinfo{note}{\url{https://github.com/gwern/mueval}}.

\bibitemdeclare{inproceedings}{clow2013moocs}
\bibitem{clow2013moocs}
\bibinfo{author}{Doug \surnamestart Clow\surnameend} (\bibinfo{year}{2013}):
  \emph{\bibinfo{title}{MOOCs and the Funnel of Participation}}.
\newblock In: {\sl \bibinfo{booktitle}{Proceedings of the Third International
  Conference on Learning Analytics and Knowledge}}, pp.
  \bibinfo{pages}{185--189}, \doi{10.1145/2460296.2460332}.

\bibitemdeclare{inproceedings}{czaplicki2013async}
\bibitem{czaplicki2013async}
\bibinfo{author}{Evan \surnamestart Czaplicki\surnameend} \&
  \bibinfo{author}{Stephen \surnamestart Chong\surnameend}
  (\bibinfo{year}{2013}): \emph{\bibinfo{title}{Asynchronous Functional
  Reactive Programming for {GUI}s}}.
\newblock In: {\sl \bibinfo{booktitle}{Proceedings of the 34th ACM SIGPLAN
  Conference on Programming Language Design and Implementation}}, pp.
  \bibinfo{pages}{411--422}, \doi{10.1145/2491956.2462161}.

\bibitemdeclare{misc}{done}
\bibitem{done}
\bibinfo{author}{Christopher \surnamestart Done\surnameend}
  (\bibinfo{year}{2014}): \emph{\bibinfo{title}{Try Haskell}}.
\newblock \bibinfo{note}{\url{https://github.com/tryhaskell/tryhaskell}}.

\bibitemdeclare{article}{findler2002drscheme}
\bibitem{findler2002drscheme}
\bibinfo{author}{Robert~Bruce \surnamestart Findler\surnameend},
  \bibinfo{author}{John \surnamestart Clements\surnameend},
  \bibinfo{author}{Cormac \surnamestart Flanagan\surnameend},
  \bibinfo{author}{Matthew \surnamestart Flatt\surnameend},
  \bibinfo{author}{Shriram \surnamestart Krishnamurthi\surnameend},
  \bibinfo{author}{Paul \surnamestart Steckler\surnameend} \&
  \bibinfo{author}{Matthias \surnamestart Felleisen\surnameend}
  (\bibinfo{year}{2002}): \emph{\bibinfo{title}{{DrScheme}: A programming
  environment for {S}cheme}}.
\newblock {\sl \bibinfo{journal}{Journal of Functional Programming}}
  \bibinfo{volume}{12}(\bibinfo{number}{2}), pp. \bibinfo{pages}{159--182},
  \doi{10.1017/S0956796801004208}.

\bibitemdeclare{misc}{type}
\bibitem{type}
\bibinfo{author}{\surnamestart {GHC Wiki}\surnameend} (\bibinfo{year}{2015}):
  \emph{\bibinfo{title}{Custom Type Errors}}.
\newblock
  \bibinfo{note}{\url{https://ghc.haskell.org/trac/ghc/wiki/Proposal/CustomTypeErrors}}.

\bibitemdeclare{misc}{ihaskell}
\bibitem{ihaskell}
\bibinfo{author}{Andrew \surnamestart Gibiansky\surnameend}
  (\bibinfo{year}{2017}): \emph{\bibinfo{title}{A Haskell kernel for IPython}}.

\bibitemdeclare{inproceedings}{guo2014how}
\bibitem{guo2014how}
\bibinfo{author}{Philip~J. \surnamestart Guo\surnameend}, \bibinfo{author}{Juho
  \surnamestart Kim\surnameend} \& \bibinfo{author}{Rob \surnamestart
  Rubin\surnameend} (\bibinfo{year}{2014}): \emph{\bibinfo{title}{How Video
  Production Affects Student Engagement: An Empirical Study of {MOOC} Videos}}.
\newblock In: {\sl \bibinfo{booktitle}{Proceedings of the First ACM Conference
  on Learning @ Scale Conference}}, pp. \bibinfo{pages}{41--50},
  \doi{10.1145/2556325.2566239}.

\bibitemdeclare{inproceedings}{hage2007mining}
\bibitem{hage2007mining}
\bibinfo{author}{Jurriaan \surnamestart Hage\surnameend} \&
  \bibinfo{author}{Peter \surnamestart {van Keeken}\surnameend}
  (\bibinfo{year}{2007}): \emph{\bibinfo{title}{Mining {H}elium programs with
  {N}eon}}.
\newblock In: {\sl \bibinfo{booktitle}{Draft Proceedings of the 8th Symposium
  on Trends in Functional Programming}}, pp. \bibinfo{pages}{35--50}.
\newblock
  \urlprefix\url{http://www.academia.edu/download/30841452/10.1.1.81.4266.pdf}.

\bibitemdeclare{inproceedings}{heeren2003helium}
\bibitem{heeren2003helium}
\bibinfo{author}{Bastiaan \surnamestart Heeren\surnameend},
  \bibinfo{author}{Daan \surnamestart Leijen\surnameend} \&
  \bibinfo{author}{Arjan \surnamestart van IJzendoorn\surnameend}
  (\bibinfo{year}{2003}): \emph{\bibinfo{title}{Helium, for Learning Haskell}}.
\newblock In: {\sl \bibinfo{booktitle}{Proceedings of the 2003 ACM SIGPLAN
  Workshop on Haskell}}, pp. \bibinfo{pages}{62--71},
  \doi{10.1145/871895.871902}.

\bibitemdeclare{article}{hudak1992report}
\bibitem{hudak1992report}
\bibinfo{author}{Paul \surnamestart Hudak\surnameend}, \bibinfo{author}{Simon
  \surnamestart Peyton~Jones\surnameend}, \bibinfo{author}{Philip \surnamestart
  Wadler\surnameend}, \bibinfo{author}{Brian \surnamestart Boutel\surnameend},
  \bibinfo{author}{Jon \surnamestart Fairbairn\surnameend},
  \bibinfo{author}{Joseph \surnamestart Fasel\surnameend},
  \bibinfo{author}{Mar{\'\i}a~M \surnamestart Guzm{\'a}n\surnameend},
  \bibinfo{author}{Kevin \surnamestart Hammond\surnameend},
  \bibinfo{author}{John \surnamestart Hughes\surnameend},
  \bibinfo{author}{Thomas \surnamestart Johnsson\surnameend} et~al.
  (\bibinfo{year}{1992}): \emph{\bibinfo{title}{Report on the programming
  language Haskell: a non-strict, purely functional language version 1.2}}.
\newblock {\sl \bibinfo{journal}{ACM SIGPLAN Notices}}
  \bibinfo{volume}{27}(\bibinfo{number}{5}), pp. \bibinfo{pages}{1--164},
  \doi{10.1145/130697.130699}.

\bibitemdeclare{article}{jordan2014initial}
\bibitem{jordan2014initial}
\bibinfo{author}{Katy \surnamestart Jordan\surnameend} (\bibinfo{year}{2014}):
  \emph{\bibinfo{title}{Initial trends in enrolment and completion of massive
  open online courses}}.
\newblock {\sl \bibinfo{journal}{International Review of Research in Open and
  Distance Learning}} \bibinfo{volume}{15}(\bibinfo{number}{1}), pp.
  \bibinfo{pages}{133--160}, \doi{10.19173/irrodl.v15i1.1651}.

\bibitemdeclare{misc}{edxmooc}
\bibitem{edxmooc}
\bibinfo{author}{Katy \surnamestart Jordan\surnameend} (\bibinfo{year}{2015}):
  \emph{\bibinfo{title}{{MOOC} Completion Rates: The Data}}.
\newblock \bibinfo{note}{\url{http://www.katyjordan.com/MOOCproject.html}}.

\bibitemdeclare{inproceedings}{lammel2013101}
\bibitem{lammel2013101}
\bibinfo{author}{Ralf \surnamestart L\"{a}mmel\surnameend},
  \bibinfo{author}{Thomas \surnamestart Schmorleiz\surnameend} \&
  \bibinfo{author}{Andrei \surnamestart Varanovich\surnameend}
  (\bibinfo{year}{2013}): \emph{\bibinfo{title}{The 101Haskell Chrestomathy: A
  Whole Bunch of Learnable Lambdas}}.
\newblock In: {\sl \bibinfo{booktitle}{Proceedings of the 25th Symposium on
  Implementation and Application of Functional Languages}}, pp.
  \bibinfo{pages}{25:25--25:36}, \doi{10.1145/2620678.2620681}.

\bibitemdeclare{inproceedings}{marceau2011measuring}
\bibitem{marceau2011measuring}
\bibinfo{author}{Guillaume \surnamestart Marceau\surnameend},
  \bibinfo{author}{Kathi \surnamestart Fisler\surnameend} \&
  \bibinfo{author}{Shriram \surnamestart Krishnamurthi\surnameend}
  (\bibinfo{year}{2011}): \emph{\bibinfo{title}{Measuring the Effectiveness of
  Error Messages Designed for Novice Programmers}}.
\newblock In: {\sl \bibinfo{booktitle}{Proceedings of the 42nd ACM Technical
  Symposium on Computer Science Education}}, pp. \bibinfo{pages}{499--504},
  \doi{10.1145/1953163.1953308}.

\bibitemdeclare{misc}{scaladata}
\bibitem{scaladata}
\bibinfo{author}{Heather \surnamestart Miller\surnameend}
  (\bibinfo{year}{2012}): \emph{\bibinfo{title}{Visualize statistics from the
  {MOOC} ``{F}unctional Programming Principles in {S}cala'' using {S}cala!}}
\newblock \bibinfo{note}{\url{https://github.com/heathermiller/progfun-stats}}.

\bibitemdeclare{misc}{scalamooc}
\bibitem{scalamooc}
\bibinfo{author}{Heather \surnamestart Miller\surnameend} \&
  \bibinfo{author}{Martin \surnamestart Odersky\surnameend}
  (\bibinfo{year}{2012}): \emph{\bibinfo{title}{Functional Programming
  Principles in Scala: Impressions and Statistics}}.
\newblock
  \bibinfo{note}{\url{http://docs.scala-lang.org/news/functional-programming-principles-in-scala-impressions-and-statistics.html}}.

\bibitemdeclare{misc}{xkcd}
\bibitem{xkcd}
\bibinfo{author}{Randall \surnamestart Munroe\surnameend}
  (\bibinfo{year}{2014}): \emph{\bibinfo{title}{Haskell}}.
\newblock \bibinfo{note}{\url{https://xkcd.com/1312/}}.

\bibitemdeclare{article}{murphy17analysis}
\bibitem{murphy17analysis}
\bibinfo{author}{Ellen \surnamestart Murphy\surnameend}, \bibinfo{author}{Tom
  \surnamestart Crick\surnameend} \& \bibinfo{author}{James~H. \surnamestart
  Davenport\surnameend} (\bibinfo{year}{2017}): \emph{\bibinfo{title}{An
  Analysis of Introductory Programming Courses at UK Universities}}.
\newblock {\sl \bibinfo{journal}{The Art, Science, and Engineering of
  Programming}} \bibinfo{volume}{1}(\bibinfo{number}{2}),
  p.~\bibinfo{pages}{18}, \doi{10.22152/programming-journal.org/2017/1/18}.

\bibitemdeclare{misc}{peytonjones2017escape}
\bibitem{peytonjones2017escape}
\bibinfo{author}{Simon \surnamestart {Peyton Jones}\surnameend}
  (\bibinfo{year}{2017}): \emph{\bibinfo{title}{Escape from the Ivory Tower:
  the Haskell Journey}}.
\newblock \bibinfo{note}{Video--watch from 15:30,
  \url{https://www.youtube.com/watch?v=re96UgMk6GQ}}.

\bibitemdeclare{misc}{logs}
\bibitem{logs}
\bibinfo{author}{Jeremy \surnamestart Singer\surnameend}
  (\bibinfo{year}{2018}): \emph{\bibinfo{title}{Anonymised Logfiles from
  TryHaskell Servers}}.
\newblock
  \bibinfo{note}{\url{https://github.com/jeremysinger/haskellmooc_logfiles}}.

\bibitemdeclare{article}{stefik2013empirical}
\bibitem{stefik2013empirical}
\bibinfo{author}{Andreas \surnamestart Stefik\surnameend} \&
  \bibinfo{author}{Susanna \surnamestart Siebert\surnameend}
  (\bibinfo{year}{2013}): \emph{\bibinfo{title}{An Empirical Investigation into
  Programming Language Syntax}}.
\newblock {\sl \bibinfo{journal}{ACM Transactions on Computing Education}}
  \bibinfo{volume}{13}(\bibinfo{number}{4}), pp. \bibinfo{pages}{19:1--19:40},
  \doi{10.1145/2534973}.

\bibitemdeclare{misc}{hugs}
\bibitem{hugs}
\bibinfo{author}{Simon \surnamestart Thompson\surnameend}
  (\bibinfo{year}{1999}): \emph{\bibinfo{title}{Some Common (and not so
  common!) Hugs Errors}}.
\newblock
  \bibinfo{note}{\url{https://www.cs.kent.ac.uk/people/staff/sjt/craft2e/errors/allErrors.html}}.

\bibitemdeclare{misc}{forkedth}
\bibitem{forkedth}
\bibinfo{author}{Wim \surnamestart Vanderbauwhede\surnameend} \&
  \bibinfo{author}{Jeremy \surnamestart Singer\surnameend}
  (\bibinfo{year}{2016}): \emph{\bibinfo{title}{Haskell Tutorials}}.
\newblock
  \bibinfo{note}{\url{https://github.com/wimvanderbauwhede/haskelltutorials}}.

\bibitemdeclare{inproceedings}{vihavainen2012multi}
\bibitem{vihavainen2012multi}
\bibinfo{author}{Arto \surnamestart Vihavainen\surnameend},
  \bibinfo{author}{Matti \surnamestart Luukkainen\surnameend} \&
  \bibinfo{author}{Jaakko \surnamestart Kurhila\surnameend}
  (\bibinfo{year}{2012}): \emph{\bibinfo{title}{Multi-faceted support for
  {MOOC} in programming}}.
\newblock In: {\sl \bibinfo{booktitle}{Proceedings of the 13th annual
  conference on {I}nformation {T}echnology {E}ducation}}, pp.
  \bibinfo{pages}{171--176}, \doi{10.1145/2380552.2380603}.

\end{thebibliography}



\end{document}